\magnification \magstep1
\raggedbottom
\openup 4\jot 
\voffset6truemm
\leftline {\bf A SPINORIAL PERSPECTIVE ON MASSLESS PHOTONS}
\vskip 0.3cm
\leftline {\bf Giampiero Esposito}
\vskip 0.3cm
\noindent
{\it INFN, Sezione di Napoli,
Complesso Universitario di Monte S. Angelo, Via Cintia, Edificio 6,
80126 Napoli, Italy; e-mail: giampiero.esposito@na.infn.it}
\vskip 0.3cm
\noindent
{\it Dipartimento di Scienze Fisiche, 
Universit\`a di Napoli ``Federico II'',
Complesso Universitario di Monte S. Angelo, Via Cintia, Edificio 6,
80126 Napoli, Italy}
\vskip 1cm
\noindent
{\bf Abstract}. We exploit the fact that, in Minkowski space-time,
$\gamma$-matrices are possibly more fundamental than the metric to
describe how gauge invariance at perturbative level enforces a
Lagrangian for spinor electrodynamics with massless photons. The term
quadratic in the potential arises naturally in the gauge-fixed
Lagrangian but has vanishing coefficient.
\vskip 100cm
\noindent
Our work in Ref. 1 has suggested a spinorial perspective on the massless
nature of photons, pointing out that the gauge-fixing functional 
$\Phi(A)$ of quantum electrodynamics (hereafter QED) can be generated by
a family of $4 \times 4$ matrices $\Phi_{i}^{\; j}(A)$ which are linear in
the potential $A_{\mu}$, according to Eq. (3.4) in Ref. 1. A better
understanding is here gained by bearing in mind that any $4 \times 4$ matrix
can be expressed in the form$^{2}$ 
$$
M_{i}^{\; j}=a \delta_{i}^{\; j}+b_{\mu}(\gamma^{\mu})_{i}^{\; j}
+{1\over 2}c_{\mu \nu}(\zeta^{\mu \nu})_{i}^{\; j}
+d_{\mu}(\zeta^{\mu})_{i}^{\; j}+e (\gamma^{5})_{i}^{\; j},
\eqno (1)
$$
where
$$
(\zeta^{\mu \nu})_{i}^{\; j} \equiv 
{1\over 2} \Bigr(\gamma^{\mu} \gamma^{\nu}
-\gamma^{\nu} \gamma^{\mu} \Bigr)_{i}^{\; j}
={1\over 2}\Bigr[(\gamma^{\mu})_{i}^{\; k} (\gamma^{\nu})_{k}^{\; j}
-(\gamma^{\nu})_{i}^{\; k} (\gamma^{\mu})_{k}^{\; j} \Bigr],
\eqno (2)
$$
$$
(\zeta^{\mu})_{i}^{\; j} \equiv (\gamma^{5} \gamma^{\mu})_{i}^{\; j}
=(\gamma^{5})_{i}^{\; l} (\gamma^{\mu})_{l}^{\; j}.
\eqno (3)
$$
We now remark that, in Eq. (1), only $\gamma^{\mu}$ and $\zeta^{\mu}$ 
contain just one space-time index and are therefore suitable for building
a $4 \times 4$ matrix $\Phi_{i}^{\; j}(A)$ linear in the potential
$A_{\mu}$. Hence we replace Eq. (3.1) in Ref. 1 by
$$
\Phi_{i}^{\; j}(A) \equiv \Bigr(\delta_{i}^{\; j}\partial^{\mu}
+\beta_{1}(\gamma^{\mu})_{i}^{\; j}
+\beta_{2}(\zeta^{\mu})_{i}^{\; j}\Bigr)A_{\mu}(x),
\eqno (4)
$$
where, for the time being, $\beta_{1}$ and $\beta_{2}$ are unknown parameters
of dimension $({\rm length})^{-1}$. We now follow our work in Ref. 1,
obtaining $\Phi(A)$ as the square root of (hereafter the space-time dimension
is taken equal to $4$)
$$ \eqalignno{
\Phi_{i}^{\; j}(A) {1\over 4} \delta_{j}^{\; k} \Phi_{k}^{\; i}(A)
&={1\over 4}\Bigr[{\rm tr}(I)(\partial^{\mu}A_{\mu})^{2}
+\beta_{1}^{2}{\rm tr}(\gamma^{\mu}\gamma^{\nu})A_{\mu}A_{\nu}
+\beta_{2}^{2}{\rm tr}(\gamma^{5}\gamma^{\mu}\gamma^{5}\gamma^{\nu})
A_{\mu}A_{\nu}\Bigr] \cr
&=(\partial^{\mu}A_{\mu})^{2}
+(\beta_{1}^{2}-\beta_{2}^{2})A_{\mu}A^{\mu},
&(5) \cr}
$$
whe\-re va\-ri\-ous ter\-ms not writ\-ten ex\-pli\-ci\-tly 
in Eq. (5) va\-ni\-sh by vir\-tue of
the fa\-mi\-li\-ar tra\-ce pro\-per\-ti\-es 
$$
{\rm tr}(\gamma^{\mu})=0, \; {\rm tr}(\gamma^{5}\gamma^{\mu})=0, \;
{\rm tr}(\gamma^{\mu}\gamma^{\nu}\gamma^{5})=0,
\eqno (6)
$$
while
$$
{\rm tr}(\gamma^{\mu}\gamma^{\nu})=4 g^{\mu \nu}, \;
{\rm tr}(\gamma^{5}\gamma^{\mu}\gamma^{5}\gamma^{\nu})
=-{\rm tr}(\gamma^{\mu}\gamma^{5}\gamma^{5}\gamma^{\nu})
=-4 g^{\mu \nu}.
\eqno (7)
$$

Furthermore, following Sec. 5 of Ref. 1, we assume multiplicative 
renormalizability, requiring that (the subscript $B$ being used for
bare quantities) 
$$
(A_{\mu})_{B}=\sqrt{z_{A}} \; A_{\mu}, \; 
\alpha_{B}={z_{A}\over z_{\alpha}}\alpha, \;
(\beta_{1})_{B}=\rho_{1} \beta_{1}, \;
(\beta_{2})_{B}=\rho_{2} \beta_{2}.
\eqno (8)
$$
The resulting physical Lagrangian density for spinor electrodynamics 
reads as (cf. Ref. 1)
$$ \eqalignno{
{\cal L}_{\rm ph}&=-{1\over 4}F_{\mu \nu}F^{\mu \nu}
+{\overline \psi}{\rm i}\gamma^{\mu}\partial_{\mu}\psi
-e {\overline \psi}\gamma^{\mu}A_{\mu}\psi -m {\overline \psi} \psi \cr
&-{1\over 2\alpha}(\partial^{\mu}A_{\mu})^{2}
-{(\beta_{1}^{2}-\beta_{2}^{2})\over 2\alpha}A_{\mu}A^{\mu},
&(9)\cr}
$$
while the part involving counterterms is given by (cf. Ref. 2)
$$ \eqalignno{
{\cal L}_{\rm ct}&=-{1\over 4}(z_{A}-1)F_{\mu \nu}F^{\mu \nu}
+(z_{\psi}-1){\overline \psi}{\rm i}\gamma^{\mu}\partial_{\mu}\psi
-(z_{e}-1)e {\overline \psi} \gamma^{\mu}A_{\mu}\psi \cr
&-(z_{m}-1)m {\overline \psi} \psi 
-{1\over 2 \alpha}(z_{\alpha}-1)(\partial^{\mu}A_{\mu})^{2} \cr
&-{1\over 2\alpha}\left[\Bigr(\rho_{1}^{2}z_{\alpha}-1)\beta_{1}^{2}
-(\rho_{2}^{2}z_{\alpha}-1)\beta_{2}^{2}\right]A_{\mu}A^{\mu}.
&(10)\cr}
$$
On requiring that $A_{\mu}A^{\mu}$ should have vanishing weight in the
counterterm Lagrangian, which is compelling to ensure that gauge
invariance is not broken by the quantum dynamics of QED, we find
$$
\rho_{1}=\pm {1\over \sqrt{z_{\alpha}}}, \;
\rho_{2}=\pm {1\over \sqrt{z_{\alpha}}}.
\eqno (11)
$$
This shows, by virtue of the last two relations in Eq. (8), that
$\beta_{1}$ and $\beta_{2}$ themselves can be identified up to a sign, i.e.
$$
\beta_{2}=\pm \beta_{1}=\pm \beta.
\eqno (12)
$$
Thus, eventually, also the physical Lagrangian is massless, and 
{\it infinitely many values} of the parameter 
$\beta$ are compatible with a massless photon.
The term $A_{\mu}A^{\mu}$, which is compatible with Lorentz and charge
conjugation invariance, exists but has vanishing weight, to preserve
gauge invariance when perturbative renormalization of multiplicative
type is implemented. Its potential occurrence is better appreciated by
virtue of spinorial geometry, since the Minkowski metric $g^{\mu \nu}$
is recovered from the $\gamma$-matrices (see Eq. (7)), and the gauge-fixing
functional $\Phi(A)$ is recovered from the $4 \times 4$ matrices
$$
\Phi_{i}^{\; j}(A) \equiv \left[\delta_{i}^{\; j}\partial^{\mu}
+\beta ((I \pm \gamma^{5})\gamma^{\mu})_{i}^{\; j}\right]A_{\mu}(x),
\eqno (13)
$$
according to$^{1}$
$$
\Phi(A)=\sqrt{\Phi_{i}^{\; j}(A) {1\over 4} \delta_{j}^{\; k}
\Phi_{k}^{\; i}(A)}.
\eqno (14)
$$
Had we not payed attention to spinorial geometry, we would have been 
unable to deal with $A_{\mu}A^{\mu}$ terms, apart from adding them by hand
to the original Lagrangian, a procedure which regrettably spoils gauge 
invariance from the very beginning. Upon imposing Eq. (12) ({\it and only at
that stage}), our approach yields eventually spinor electrodynamics in
the Lorenz$^{3}$ gauge.
\vskip 100cm
\leftline {\bf REFERENCES}
\vskip 1cm
\noindent
\item {1.}
G. Esposito, ``On the occurrence of mass in field theory'', 
{\it Found. Phys.} {\bf 32}, 1459 (2002).
\item {2.}
B. S. DeWitt, {\it Dynamical Theory of Groups and Fields}
(Gordon \& Breach, New York, 1965).
\item {3.}
L. Lorenz, ``On the identity of the vibrations of light with electrical
currents'', {\it Phil. Mag.} {\bf 34}, 287 (1867).
\bye